\documentclass{ws-procs9x6}
\begin{document}

\def\AJ{{\it Astron. J.} }
\def\ARAA{{\it Annual Rev. of Astron. \& Astrophys.} }
\def\ApJ{{\it Astrophys. J.} }
\def\ApJL{{\it Astrophys. J. Letters} }
\def\ApJS{{\it Astrophys. J. Suppl.} }
\def\ApP{{\it Astropart. Phys.} }
\def\AA{{\it Astron. \& Astroph.} }
\def\AAR{{\it Astron. \& Astroph. Rev.} }
\def\AAL{{\it Astron. \& Astroph. Letters} }
\def\JGR{{\it Journ. of Geophys. Res.}}
\def\JPhG{{\it Journ. of Physics} {\bf G} }
\def\PhFl{{\it Phys. of Fluids} }
\def\PR{{\it Phys. Rev.} }
\def\PRD{{\it Phys. Rev.} {\bf D} }
\def\PRL{{\it Phys. Rev. Letters} }
\def\Nature{{\it Nature} }
\def\MNRAS{{\it Month. Not. Roy. Astr. Soc.} }
\def\ZA{{\it Zeitschr. f{\"u}r Astrophys.} }
\def\ZFN{{\it Zeitschr. f{\"u}r Naturforsch.} }
\def\etal{{\it et al.}}
%
\def\simle{\lower 2pt \hbox {$\buildrel < \over {\scriptstyle \sim }$}}
\def\simge{\lower 2pt \hbox {$\buildrel > \over {\scriptstyle \sim }$}}


\title{ Origin and physics of the highest energy cosmic rays: What can we 
learn from Radio Astronomy ?}

\author{Peter L. Biermann\footnote{E-mail:plbiermann@mpifr-bonn.mpg.de}$^{1,2,3}$, P. Gina  Isar $^4$, Ioana C. Maris$^4$, 
Faustin Munyaneza$^1$, \& Oana Ta\c{s}c\u{a}u$^5$}
\address{
$^1$Max-Planck Institute for Radioastronomy, Bonn, Germany\\
$^2$Department of Physics and Astronomy,
University of Bonn, Bonn, Germany, \\
$^3$Department of Physics and Astronomy, University of Alabama,
Tuscaloosa, AL, USA\\
$^4$ FZ Karlsruhe, and Physics Dept., Univ. of Karlsruhe \\
$^5$ University of Wuppertal, Wuppertal, Germany}


\begin{abstract}

Here in this lecture we will touch on two aspects, one the new radio 
methods to observe the effects of high energy particles, and second the 
role that radio galaxies play in helping us understand high energy 
cosmic rays.  We will focus here on the second topic, and just review 
the latest developments in the first.  Radio measurements of the 
geosynchrotron radiation produced by high energy cosmic ray particles 
entering the atmosphere of the Earth as well as radio \v{C}erenkov 
radiation coming from interactions in the Moon are another path; radio 
observations of interactions in ice at the horizon in Antarctica is a 
related attempt.  Radio galaxy hot spots are prime candidates to produce 
the highest energy cosmic rays, and the corresponding shock waves in 
relativistic jets emanating from nearly all black holes observed.  We 
will review the arguments and the way to verify the ensuing 
predictions.  This involves the definition of reliable samples of active 
sources, such as black holes, and galaxies active in star formation.  
The AUGER array will probably decide within the next few years, where 
the highest energy cosmic rays come from, and so frame the next quests, 
on very high energy neutrinos and perhaps other particles.
\end{abstract}

\keywords{Cosmic rays, magnetic fields, active galactic nuclei, black holes, radio galaxies}
\bodymatter

\section{Introduction}

Recent years have seen a proliferation in new experimental efforts to 
measure very high energy particles, both in actually constructing huge 
new arrays like AUGER in Argentina, but also new attempts to measure 
high energy particles in new ways, mostly focussing on the radio range.  
We cannot but mention in passing the hugely successful telescopes for 
TeV gamma rays, like HESS, MAGIC, MILAGRO and Cangaroo, as well as the 
neutrino observatory IceCube for high energy neutrinos.  Here we focus 
on yet higher energies, near and beyond EeV (= $10^{18}$ eV). For the subject covered in this 
lecture, the reader is refered to important 
textbooks \cite{berezinskii90, gaisser90,stanev04}. 

On the other hand, our theoretical understanding of possible production 
of high energy charged particles, neutrinos and photons is also reaching 
a measure of maturity, with many efforts concentrating on the role of 
shock waves and flows in relativistic jets, such as in radio galaxy hot 
spots as discussed in various review articles \cite{biermann94, halzenhooper02,
 piran99, sigl01,wieberbiermann99}.
 
\section{Radio detection methods}

It has been recognized many decades ago, that high energy particles 
produce secondary radio emission. The emission is now far along to be 
understood as geosynchrotron emission \cite{falckegorham03,falcke04,falcke05},
 or as radio \v{C}erenkov emission.  Now we have 
established efforts under way to observe, calibrate and use such 
emissions to set limits, possibly soon measure, high energy neutrinos, 
very high energy cosmic rays and also unknown particles.

The furthest along has been the effort to use geosynchrotron emission, 
when the airshower is a directly visible radio spot in the sky.  The 
observation of high energy cosmic rays will be incorporated into the 
LOFAR array, in the Netherlands; the LOPES array in Karlsruhe, Germany, 
undertakes the control, and calibration of these
 emissions \cite{apel06, gemmeke06, haungs06a,haungs06b,horandel06,
 horneffer06,huegefalcke05,huege06,nehls06,petrovics06}.
  Radio \v{C}erenkov emission 
from the Moon is another effort, as is the corresponding observation in 
ice at the horizon in Antarctica \cite{barwick06a,barwick06b,saltzberg06}.
   There are corresponding efforts elsewhere, and also some 
tests have been done to use salt-domes~\cite{gorham02}.

\section{Active Galactic Nuclei}

For Galactic cosmic rays clearly supernova explosions are the prime 
candidates to be sources of cosmic rays; the latest results from the TeV 
telescopes strongly support this expectation and interpretation.  The 
energy range to which cosmic rays derive from Galactic sources is not 
entirely certain, but the transition to extragalactic cosmic rays is 
somewhere near $3 \, 10^{18}$ eV, perhaps at slightly lower energy.

Where ultra high energy cosmic rays come from is not certain, but the 
most promising candidates are radio galaxy hot spots and other shocks in 
relativistic jets, and gamma ray bursts, almost certainly also a 
phenomenon involving ultra-relativistic jets.

Here we wish to concentrate on radio galaxies, as in their case the 
argument is persuasive, that they accelerate cosmic rays to extremely 
high energy.  This does not necessarily imply that they are the sources 
for the events we observe.

\section{Radio galaxies}

Radio galaxy hot spots as well as many other knots in radio jets show an 
ubiquitous cutoff near $3 \, 10^{14}$ Hz, originally discovered in the 
mid 1970ies. Radio, infrared and optical observations strongly suggest 
that these hot spots and knots are weakly relativistic shock waves.  The 
basic phenomenon seems to be quite independent of external 
circumstances, and so requires a very simple mechanism. It has been
 shown that in such shocks protons can be 
accelerated \cite{biermannstritt87}, subject to synchrotron and 
Inverse Compton losses, then 
initiate a wave-field in the turbulent plasma, and the electrons 
scattered and accelerated in the shock are then limited to emit at that 
synchrotron frequency observed, $\nu^{\star} \simeq 3 \, 10^{14}$ Hz, 
almost independent of details.

\subsection{The maximum energy}

The maximum energy of the protons $E_{p,max}$ from this loss limit 
implicated can then be written as

\begin{equation}
E_{p,max}  \simeq \, 1.5 \, 10^{20} \, {\rm eV} \, \, 
\left(\frac{\nu^{\star}_e}{ 3.
10^{14} \, {\rm Hz}}\right)^{1/2} \, B^{-1/2}
\end{equation}

Here we are independent of all the detailed assumptions about the 
intensity of the turbulence, and the exact shock speed; the dependence 
through the magnetic field on other parameters is only with the 
1/7-power; typical magnetic field strength inferred are between 
$10^{-2}$ and $10^{-4}$ Gau{\ss}.  

There is a corresponding limit from the requirement \cite{hillas84},
 that 
the Larmor motion of the particle fit into the available space; this can 
be written as $E_{p,max} \simeq 10^{21} \, {\rm eV} \, L_{46}^{1/2}$, 
where $L_{46}$ is the flow of energy along the jet, some fraction of the 
accretion power to the black hole, in units of $10^{46}$ erg/s.
Therefore radio-galaxies are confirmed source candidates for protons at 
energies $> \; 10^{20} $ eV!

\subsection{Positional correlations}

Of course it is interesting to look for associations of ultra high 
energy events and their arrival directions with known objects in the 
sky, whether it is the supergalactic 
plane \cite{devaucouleurs56,stanev95},
 some distant objects\cite{farrarbiermann98},
  or nearby 
galaxies with their black holes.  Here we report work with Ioana 
Maris, done in the years up to 2004, reported on our web-page, and 
in many lectures.

In order to have a statistical meaningful result we have to form 
complete samples of candidate sources; in some cases irregular samples 
will be unavoidable.  We need active galactic nuclei, starving and 
active, starburst and normal galaxies, clusters of galaxies, with the 
reasoning that all such candidates could produce ultra high energy 
cosmic rays, be in shocks in jets, in gamma ray bursts, in accretion 
shocks to clusters, and hyperactive other stars.  Recently merged black 
holes would be another hypothetical option, as then exotic particles 
might be produced.  Most plausible would be relativistic jets pointed at 
Earth, also known as flat spectrum radio sources; all BL Lac type 
sources are in this class, although not all flat spectrum radio sources 
are of BL Lac type.

In any new search, such as to identify sources for high energy 
neutrinos, or sources for straight propagation of particles in the AUGER 
data, one should follow the same route:  use complete well defined and 
small samples, and define them beforehand.  With the IceCube 
collaboration we have just completed this task, and it has been 
published\cite{achterberg06}.

We have done this task for ultra high energy cosmic rays some time ago 
with the data available publicly, using the set of accessible ultra high 
energy events with good directional accuracy above $4 \, 10^{19}$ eV: 80 
events: AGASA (61), Haverah Park (6), Yakutsk (12), and Fly's Eye (1);
 we  also used  one event which was 38 EeV, as this was the sample used also by
AGASA for the doublets analysis. 
  
This has been done before by many, from about 1960 by Ginzburg, 
including us from 1985,  and in recent
 papers~\cite{tinyakov01,evans03,finleywesterhoff04}.
 
We find, similar to some other searches, that radio sources from the 
Condon Radio survey in positional coincidence with far infrared (IRAS) 
sources do show a highly improbable association with ultra high energy 
cosmic ray events.  We can go one step further and use jet-disk 
symbiosis~\cite{falcke95} to predict maximum particle 
energy and maximum cosmic ray flux, and so check whether these so 
identified sources can possibly produce the flux observed, even 
closely.  In work with Heino Falcke, Sera Markoff, and Feng Yuan as well 
as in their own work these concepts about the physics of relativistic 
jets have been tested at all levels of observability, from microquasars 
via low luminosity active galactic nuclei to more powerful sources.  
Current work is being done with Marina Kaufman.  Tests include fitting 
the entire electromagnetic spectrum, and the variability.

Assuming that these particles can violate the GZK interaction with the 
microwave background, these sources could in fact account for the high 
flux at high particle energy, quite surprisingly\cite{biermann0106}.
 Also, many of the sources so identified are quite famous in 
radio astronomy history, sources such as 3C120, 3C147, 4C39.25, 3C449 
and the like.  We did identify a speculative picture in which such 
events coming from large cosmological distances could be 
explained\cite{biermannframpton06}, 
 using particles in higher dimensions and 
the distortion of space time close to the merger of two spinning black 
holes.  Should be here no such correlation once we have much more 
extended samples, it might be possible to set limits on this kind of 
physics.

However, since we tried many times to get such a result we hesitate to 
assign
any physical meaning to this physical picture at this time.

\section{Magnetic fields}

The magnetic fields filling the cosmos are generally too weak to 
influence the propagation of ultra high energy cosmic rays by more han a 
few degrees~\cite{ryu98}  There may be special environments where 
this may not be true, like the boundary regions around radio galaxies in 
a cluster of galaxies.

However, the halos and winds of individual galaxies such as ours may 
have an appreciable influence:  For starburst galaxies such as M82 the 
existence of a wind has been long shown~\cite{kronberg85,heckman87}.
  In our Galaxy this was  finally recently demonstrated \cite{westmeier05},
   for NGC1808 it seems obvious in HST pictures, and the 
magnetic nature of the wind was demonstrated in the example of NGC4569  
 through radio polarization images \cite{chyzy06}.  These 
last observations also confirm the basic ansatz~\cite{breits91} 
 which we have followed in our work
 with Lauren\c{t}iu Caramete.  The key point is that the bending in a magnetic field 
standard topology such as a Parker wind~\cite{parker58} with $B_{\phi} \sim \sin 
\theta /r$ can be large as the Lorentz force is an integral in $d r$ 
over $1/r$, where $r$ is the distance from the center, and $\theta$ is 
the zenith angle in polar coordinates.  Another more subtle point is 
that the turbulence in the wind is likely to be $k^{-2}$, a saw-tooth 
pattern, since it is caused by shock waves running through the medium 
produced by OB star bubbles and their subsequent supernova explosions.  
This is in contrast to the turbulence in the thick hot 
disk \cite{snowden97},
 where observations have shown directly, that in a 3D 
isotropic approximation the turbulence is of Kolmogorov nature, so is 
$k^{-5/3}$, where $k$ is the wavenumber.

So the general concept of our Galaxy which we use, has three components: 
A cool thin disk, with lots of neutral and molecular gas, with about 200 
pc thickness, but permeated by tunnels of hot gas~\cite{coxsmith74};
 a 
hot disk, with low density and high temperature with a thickness of 
about 4 kpc, and a magnetic wind, to a very first approximation akin to 
a Parker wind, extending possibly quite far out, such as a few hundred 
kpc, or even a Mpc (from ram pressure arguments).

In work with Alex Curu\c{t}iu we have calculated in a first 
approximation the scattering of ultra high energy cosmic rays in such an 
environment, assuming various possible sources, such as M87 or Cen A.  
We have tested different assumptions about the level of turbulence, and 
find that maximal turbulence alone would reproduce the homogeneous sky 
distribution which has been observed at 30 EeV.

The predicted sky distribution (see the 2004 report on our web-page) 
consists then of long irregular stripes across the sky, as M87 is quite 
close to the Galactic North pole.  One of these stripes is roughly in 
the same region as the supergalactic plane in the Southern and part of 
the Northern sky. This has been our prediction for some years now.

\subsection{Nearby sources}

In other work with Oana Ta\c{s}c\u{a}u, Ralph Engel, Heino Falcke, Ralf 
Ulrich, and Todor Stanev we have identified all available data on nearby 
black holes, and calculated their cosmic ray maximal contribution, in 
particle energy and in flux.  For most sources the maximal energy of the 
particles is quite small, below or near EeV, and the flux is also low.  
However, for a few sources, such as M87, M84, Cen A, and NGC1068 the 
flux is of interest.  At the highest energy only M87 competes, as 
already suggested many years ago by Ginzburg~\cite{ginzburg63}, and Watson, and in 
a detailed physical model in Biermann \& Strittmatter\cite{biermannstritt87}.

The list has also been available on our web-page since 2004.

There is a question, however, already noted above, that a source may be 
strong, but how many of these particles make it out of the relativistic 
bubble (visible in  low frequency radio data) around the radio galaxy \cite{owen00}, 
and and then on towards us?  To fit the data as 
compiled in the PDG report\cite{gaisserstanev02} the flux from M87 
has to be reduced by about a factor 4, and so at 1 EeV Cen A is about 20 
times stronger than M87, but does not itself extend much beyond 
$10^{19}$ eV.

Then just adding the contributions from the strongest sources, and 
running them through a Monte-Carlo for propagation simulation reproduces 
the observed spectrum quite well.

\subsubsection{Samples for testing}

Above we have used those active black holes, for which we have data. But 
other approaches are also possible and would need to be tested.  Such 
sample definitions have recently been developed in collaboration with 
the IceCube collaboration~\cite{achterberg06}.  Here the sample 
selection has two key differences:  First, ultra high energy cosmic rays 
are expected to be protons, so they interact with the microwave 
background, and so almost certainly come from nearby in the cosmos, 
about 50 Mpc or less. Second, protons are charged and so they may 
deviate from a straight line in their propagation. Here we focus on the 
sources.

Even among the nearby sources, there are many possible candidates, one
could consider:
\begin{itemize}
\item{}  Black holes are quite common, and so almost all galaxies have a 
massive black hole\cite{faber97}.
  Usually the activity of such a 
black hole is very limited, but experience demonstrates that it is 
almost always detectable in radio emission, which we interpret as the 
emission from a relativistic jet \cite{chini89}.
  The 
accretion rate to power this jet could derive from just the wind of a 
neighboring red giant star.  Also, the data suggest that black holes in 
the centers of galaxies always have a mass larger than about $10^5$ or 
$10^6$ solar masses \cite{greene06}; we have argued that this 
minimum mass derives from the first growth of a dark matter 
star \cite{munyabiermann05,munyabiermann06},
possible if in a merger of two 
galaxies the central dark matter density diverges, builds up a 
degenerate configuration, a dark matter star, which can then be eaten by 
a stellar black hole.  Dark matter accretion has no Eddington limit, 
only an angular momentum transport limit, and that should be efficient 
during the highly disturbed situation of a merger of two galaxies.  On 
the other hand, there might be some subtlety about black hole physics, 
that we are missing, and so we should obviously just check.  The largest 
challenge to our physical understanding will appear if we find a 
correlation with very low mass black holes and ultra high energy 
events.  At first, however, this implies just a tally of black holes in 
our cosmic neighborhood, ordered by mass.  And as the mass of the black 
hole directly relates to the mass of the spheroidal population of stars, 
or the bulge, as well as the stellar central velocity dispersion, we 
have to start the search with these properties of galaxies.

\item{}  Active black holes, such as M87, NGC315, NGC5128 (= Cen A), all 
have relativistic jets, and can accelerate particles.  Since the power 
and also the maximum article energy scale with the radio emission, one 
needs a sample of black holes with measured compact emission, and so we 
need a sample ordered by predicted cosmic ray flux at energies beyond 
about a few EeV.  This is what we attempted above.

\item{}  If wind supernovae such as exploding Wolf Rayet stars manage to 
accelerate particles to 3 EeV, which happens to be the cutoff energy 
also for Galactic confinement, why not  even further?  These stars do 
not know about the Galaxy.  And if some of these stars under unknown 
special circumstances produce Gamma Ray Bursts, the particle energy 
might also be much higher.  So this implies a sample of galaxies strong 
in the far infrared, where strong star formation galaxies, or starburst 
galaxies, are very prominent.  If the flux in cosmic rays scales 
monotonically with the star formation rate, then a sample of galaxies 
ordered by flux density at 60 micron would be the best sample to test.

\item{}  If the propagation is delayed only slightly in its interaction 
with the intergalactic magnetic field, then also a direct connection 
with recent Gamma Ray Bursts might be worth investigating.

\item{}  Accretion shocks to clusters of galaxies can also accelerate 
particles\cite{kang97}, although probably not to 100 EeV; 
however, we should actually check with observations.  Most clusters, 
however, are very distant, and such a concept would qualify best for the 
Virgo cluster, again identifying just one most likely object in the sky, 
just like the radio galaxy M87, which is one of the dominant galaxies in 
the Virgo cluster.

\end{itemize}

\section{Predictions}

We have presented a theory to account for the entire cosmic ray
spectrum beyond the GZK cutoff.  This proposal is based on a physical 
and tested model for relativistic jets.

There is a Galactic magnetic wind, driven by the normal cosmic
rays.  This wind extends to some fraction of a Mpc.  The existence of a 
Galactic wind in our Galaxy is now confirmed~\cite{westmeier05}.
  The basic magnetic field topology is probably of Parker type.  
The turbulence in the wind is probably sawtooth pattern, i.e. $k^{-2}$, 
and its turbulence is close to maximal.  All galaxies with an 
appreciable level of star formation have such a wind, and their 
environment should look like Swiss cheese, embedded in the supergalactic 
sheets (work by L. Caramete).

The only contributor for cosmic ray particles beyond the GZK
cutoff is M87, with Cen A very strongly contributing just below that 
characteristic energy, with a small contribution from
NGC1068. Weaker sources are negligible due to their low maximum particle
energy, and also due to their small flux.

The arrival directions on the sky are smooth around 30 EeV,
and begin to become patchy at higher energies, showing some 
characteristic stripes, in a very simple Parker type model for the 
magnetic field topology of our Galactic wind.

If the arrival directions are smooth to the highest energies,
then this source model fails, and we require Lorentz Invariance
Violation\cite{amelino01}, new particles, 
topological defect or relic decay\cite{bhattasigl00},
 dark 
matter decay or possibly hints of quantum gravity.

Should the proposal be confirmed we can develop sources such 3 C147 as 
testbeds for particle physics - a CERN / Stanford / Fermilab in the sky.

\section{Acknowledgement}

P.L. Biermann would like to acknowledge Eun-Joo Ahn, Julia Becker, Venya 
Berezinsky,
Geoff Bicknell, Genadi Bisnovatyi-Kogan, Sabrina Ca\-sanova, Mihaela 
Chirvasa, Alina Donea, Ralph Engel, Torsten Ensslin, Heino Falcke, Paul 
Frampton, Cristina Galea, Laszlo Ger\-gely, Andreas Gross, Francis 
Halzen, Alejandra Kandus, Hyesung Kang, Marina Kaufman-Bernard{\'o},  
Gopal Krish\-na, Phil Kronberg, Alex Kusenko, Norbert Langer, Hyesook 
Lee, Sera Markoff, Gustavo Medina-Tanco, Athina Meli, 
Sergej Moiseenko,  Biman Nath, Angela Olinto, Adrian 
Popescu, Ray Protheroe, Giovanna Pug\-liese, Wolfgang Rhode, Sorin 
Roman, Gustavo Romero, Dongsu Ryu, Norma Sanchez, Gerd Sch{\"a}\-fer, 
Eun-Suk Seo, Maury Shapiro, Ramin Sina, Todor Stanev, Jaroslaw 
Stasielak, Samvel Ter-Anton\-yan, Valeriu Tudose, 
Ralf Ulrich, Marek Urbanik, Ana Vasile, Hector de Vega, Yiping Wang, 
Alan Watson, John Wefel, Stefan Westerhoff, Paul Wiita, Arno Witzel, 
Gaurang Yodh, Feng Yuan, Cao Zhen, Christian Zier, now T. Kellmann, I. 
Du\c{t}an, L. Caramete, A. Curu\c{t}iu, Alina Istrate, ..., ...

Special support comes from the European Union Sokrates / Erasmus grants in
collaboration with East-European Universities, with partners T. Zwitter 
(Ljubljana, Slovenia), L. Gergely (Szeged, Hungary), M. Ostrowski 
(Cracow,  Poland), K. Petrovay (Budapest, Hungary), A. Petrusel 
(Cluj-Napoca, Romania), and M.V. Rusu (Bucharest, Romania).  

Work with PLB is supported through the AUGER theory and membership grant 
05~CU~5PD1/2 via DESY/BMBF (Germany), and VIHKOS through the FZ Karlsruhe.

\end{document}